\documentclass[12pt]{revtex4}

\begin{document}

\title{Distributed Gaussian Polynomials as q-oscillator eigenfunctions}

\author{Hasan Karabulut}

\address {Rize University,\\
Faculty of Arts and Sciences, Physics department,\\
53100 Rize/TURKEY}

\begin{abstract}
Karabulut and Sibert (\textit{J. Math. Phys}. \textbf{38} (9), 4815
(1997)) have constructed an orthogonal set of functions from linear
combinations of equally spaced Gaussians. In this paper we show that
they are actually eigenfunctions of a q-oscillator in coordinate
representation. We also reinterpret the coordinate representation
example of q-oscillator given by Macfarlane as the functions
orthogonal with respect to an unusual inner product definition. It
is shown that the eigenfunctions in both q-oscillator examples are
infinitely degenerate.
\end{abstract}

\maketitle

\section{Introduction}

Distributed Gaussians are a set of equally spaced Gaussians: $g_{n}(x)=%
\mathrm{e}^{-c^{2}(x-n)^{2}}$ where $(n=-\infty ,...\infty ).$ A\ finite
chain of them is often used in variational calculations as a basis set. They
are a very flexible and efficient basis set often yielding very accurate
variational results. Calculating the potential matrix elements are often the
most difficult part of a variational calculation. Because of their
compactness it is very easy to calculate potential energy matrix elements
with a few point Gauss-Hermite quadrature very accurately which is a major
advantage of using this basis.

We need orthogonal functions for variational calculations and the
distributed Gaussians are not orthogonal. In a 1997 paper Karabulut and
Sibert\cite{JMP97} constructed a set of orthogonal functions from
distributed Gaussians and they studied the underlying Gaussian quadrature.
They called these functions distributed Gaussian polynomials (DG\
polynomials briefly). Their motivation was entirely practical and they were
not looking for an algebraic structure behind these function. Later Karabulut%
\cite{JMP05} showed how to normalize them and used these functions to
construct a Wannier function set from distributed Gaussians.

While searching for an operator that admits the DG polynomials as
eigenfunctions the author came across a seminal paper by Macfarlane\cite
{Macfarlane} in which he constructed a coordinate representation of his
q-oscillator algebra. (the same q-oscillator was also studied by Biederharn%
\cite{Biedenharn} but he did not have the explicit coordinate representation
example that Macfarlene gave). Eigenfunctions of Macfarlene were a linear
combination of distributed Gaussians and it involved the q-binomial
coefficients just like the DG polynomials. They looked similar to the DG
polynomials but they were not the same. Following his example, the author
constructed another coordinate representation of the q-oscillator algebra
that yields the DG polynomials as eigenfunctions. The q-oscillator turns out
to be a coordinate representation example of the Arik-Coon oscillator\cite
{Arik}. Macfarlane gave an inner product definition for his functions in
term of Rogers-Szeg\"{o} polynomials. I also found a simpler inner product
definition for his functions and reinterpreted his results.

The outline of the paper is the following. In the second section I summarize
basic results about the DG polynomials and discuss its links to Rogers-Szeg%
\"{o} polynomials. In the third section I derive the DG polynomials from the
q-oscillator algebra. In the fourth section I discuss Macfarlane's example
to his q-oscillator and reinterpret its eigenfunctions. Finally, in the last
section I give a summary and discussion.

\section{Distributed Gaussian Polynomials and their properties}

We will mostly denote the Gaussians $\mathrm{e}^{-c^{2}x^{2}}$ as $q^{x^{2}}$
where $q=\mathrm{e}^{-c^{2}}$ and $c^{-1}$ is related to the width of the
Gaussians. DG polynomials are defined as
\begin{equation}
\Phi _{n}(x)=\sum_{k=0}^{n}C_{k}^{n}(-1)^{k}q^{-k/2}q^{(x-k)^{2}}.
\label{a10}
\end{equation}
The $C_{k}^{n}$ are the well known q-binomial coefficients
\begin{equation}
C_{k}^{n}=\frac{(q,q)_{n}}{(q,q)_{k}(q,q)_{n-k}},  \label{a20}
\end{equation}
where $(q,q)_{n}$ is defined as
\begin{equation}
(q,q)_{n}=(1-q)(1-q^{2})...(1-q^{n}),  \label{a30}
\end{equation}
and $(q,q)_{0}=1$. They satisfy the following orthogonality relation\cite
{JMP05}:
\begin{equation}
\int\limits_{-\infty }^{\infty }\Phi _{n}(x)\Phi _{m}(x)\mathrm{d}x=\left\|
\Phi _{n}(x)\right\| ^{2}\delta _{nm},  \label{a40}
\end{equation}
where the norm $\left\| \Phi _{n}(x)\right\| $ is given as
\begin{equation}
\left\| \Phi _{n}(x)\right\| =\left( \frac{\pi }{2c^{2}}\right)
^{1/4}q^{-n/2}\sqrt{(q,q)_{n}}.  \label{a50}
\end{equation}
We will denote the normalized functions with lowercase $\phi $%
\begin{equation}
\phi _{n}=\frac{\Phi _{n}(x)}{\left\| \Phi _{n}(x)\right\| }=\frac{\alpha }{%
\sqrt{(q,q)_{n}}}\sum_{k=0}^{n}C_{k}^{n}(-1)^{k}q^{(n-k)/2}q^{(x-k)^{2}},
\label{a60}
\end{equation}
where
\begin{equation}
\alpha =\left( \int\limits_{-\infty }^{\infty }q^{2x^{2}}\mathrm{d}x\right)
^{-1/2}=\left( \frac{2c^{2}}{\pi }\right) ^{1/4}.  \label{a70}
\end{equation}
We defined $\alpha $ this way for later convenience.

Karabulut and Sibert\cite{JMP97} also found that the DG\ polynomials yield
harmonic oscillator eigenfunctions in a particular limit as
\begin{equation}
\lim_{c\rightarrow 0}\frac{\Phi _{n}(s/\sqrt{2}c)}{(-c/\sqrt{2})^{n}}=%
\mathrm{e}^{-s^{2}/2}h_{n}(s),  \label{a74}
\end{equation}
where $h_{n}(s)$ are the standard Hermite polynomials. We will refer to this
limit later.

DG\ polynomials are related to the Stieltjes-Wigert polynomials. Let us
write the orthogonality relation as follows
\begin{equation}
\begin{array}{ll}
\int\limits_{-\infty }^{\infty }\Phi _{n}(x-s)\Phi _{m}(x-s)\mathrm{d}x=0, &
(n\neq m).
\end{array}
\label{a80}
\end{equation}
If we denote $u=q^{-2x}$ then $\Phi _{n}(x-s)$ is written as
\begin{equation}
\Phi _{n}(x-s)=\mathrm{e}^{-(\ln u)^{2}/(-4\ln q)}u^{s}P_{n}(u;s),
\label{a90}
\end{equation}
where the polynomials $P_{n}(u;s)$ are
\begin{equation}
P_{n}(u;s)=\sum\limits_{k=0}^{n}C_{k}^{n}(-1)^{k}q^{(k+s)^{2}-k/2}u^{k}.
\label{a100}
\end{equation}
Then the orthogonality relation becomes
\begin{equation}
\begin{array}{ll}
\int\limits_{-\infty }^{\infty }\mathrm{e}^{-(\ln u)^{2}/(-2\ln
q)}u^{2s-1}P_{n}(u;s)P_{m}(u;s)\mathrm{d}x=0, & (n\neq m).
\end{array}
\label{a110}
\end{equation}
Evidently the polynomials $P_{n}(u;s)$ are orthogonal with respect to the
weight function
\begin{equation}
W(u)=\mathrm{e}^{-(\ln u)^{2}/(-2\ln q)}u^{2s-1}.  \label{a120}
\end{equation}
For $s=1/2$ the weight function is the lognormal distribution and the
corresponding polynomials are known as the Stieltjes-Wigert polynomials. So
the $P_{n}(u;1/2)$ are proportional to Stieltjes-Wigert polynomials.

The above connection to the Stieltjes Wigert polynomials were noted in
Karabulut and Sibert\cite{JMP97}. Later Atakishiyev and Nagiyev\cite{Ata}
found a connection between the Stieltjes-Wigert polynomials and
Rogers-Szeg\"{o} polynomials through the Fourier transform which implied
that DG\ polynomials are also connected the Rogers-Szeg\"{o} polynomials.
Here we note this connection.

We take the convention for the Fourier transform as
\begin{equation}
f(\theta )=\int\limits_{-\infty }^{\infty }\mathrm{e}^{\mathrm{i}2\pi \theta
x}f(x)\mathrm{d}x.  \label{a130}
\end{equation}
Using the Parseval relation of the Fourier transforms
\begin{equation}
\int\limits_{-\infty }^{\infty }F^{*}(x)G(x)\mathrm{d}x=\int\limits_{-\infty
}^{\infty }F^{*}(\theta )G(\theta )\mathrm{d}\theta ,  \label{a140}
\end{equation}
we write the eq. (\ref{a40}) as
\begin{equation}
\int\limits_{-\infty }^{\infty }\Phi _{n}^{*}(\theta )\Phi _{m}(\theta )%
\mathrm{d}\theta =\ \left( \frac{\pi }{2c^{2}}\right)
^{1/2}q^{-n}(q,q)_{n}\delta _{nm},  \label{a150}
\end{equation}
where $\Phi _{n}(\theta )$ is the Fourier transform of $\Phi _{n}(x)$ given
as
\begin{equation}
\Phi _{n}(\theta )=\left( \frac{\pi }{c^{2}}\right) ^{1/2}\mathrm{e}^{-(\pi
/c)^{2}\theta ^{2}}\sum_{k=0}^{n}C_{k}^{n}(-q^{-1/2}e^{i2\pi \theta })^{k}.
\label{a160}
\end{equation}
The polynomials
\begin{equation}
H_{n}(x)=\sum\limits_{k=0}^{n}C_{k}^{n}x^{k}  \label{a170}
\end{equation}
are known as Rogers-Szeg\"{o} polynomials\cite{Ata,Szego,Andrews}. Using
them the orthogonality relation is written as
\begin{equation}
\int\limits_{-\infty }^{\infty }H_{n}(-q^{-1/2}\mathrm{e}^{-\mathrm{i}2\pi
\theta })H_{n}(-q^{-1/2}\mathrm{e}^{\mathrm{i}2\pi \theta })\mathrm{e}%
^{-2(\pi /c)^{2}\theta ^{2}}\mathrm{d}\theta =\left( \frac{c^{2}}{2\pi }%
\right) ^{1/2}q^{-n}(q,q)_{n}\delta _{nm}.  \label{a180}
\end{equation}
A form of the Poisson summation formula reads
\begin{equation}
\int\limits_{-\infty }^{\infty }f(x)\mathrm{d}x=\int\limits_{0}^{1}\left(
\sum\limits_{k=-\infty }^{\infty }f(x+k)\right) \mathrm{d}x,  \label{a190}
\end{equation}
which is valid when $\sum\limits_{k=-\infty }^{\infty }f(x+k)$ exists (in
our case it does). Using this eq. (\ref{a180}) is expressed as
\begin{equation}
\int\limits_{0}^{1}H_{n}(-q^{-1/2}\mathrm{e}^{-\mathrm{i}2\pi \theta
})H_{n}(-q^{-1/2}\mathrm{e}^{\mathrm{i}2\pi \theta })\left(
\sum\limits_{k=-\infty }^{\infty }\mathrm{e}^{-2(\pi /c)^{2}(\theta
+k)^{2}}\right) \mathrm{d}\theta =\left( \frac{c^{2}}{2\pi }\right)
^{1/2}q^{-n}(q,q)_{n}\delta _{nm}.  \label{a200}
\end{equation}
The sum in parenthesis is periodic with period unity and it is a form of
theta function. We can expand it in Fourier series. The Fourier coefficients
can be calculated using the Poisson summation formula as
\begin{equation}
\sum\limits_{k=-\infty }^{\infty }\mathrm{e}^{-2(\pi /c)^{2}(\theta +k)^{2}}=%
\sqrt{\frac{c^{2}}{2\pi }}\sum\limits_{k=-\infty }^{\infty }q^{n^{2}/2}%
\mathrm{e}^{\mathrm{i}2\pi n\theta }.  \label{a210}
\end{equation}
Using the Jacobi $\vartheta _{3}$ function\cite{Whittaker} defined as
\begin{equation}
\vartheta _{3}(\theta ;q)=\sum\limits_{n=-\infty }^{\infty }q^{n^{2}/2}%
\mathrm{e}^{\mathrm{i}n\theta },  \label{a220}
\end{equation}
the orthogonality is written as follows:
\begin{equation}
\int\limits_{0}^{1}H_{n}(-q^{-1/2}\mathrm{e}^{-\mathrm{i}2\pi \theta
})H_{m}(-q^{-1/2}\mathrm{e}^{\mathrm{i}2\pi \theta })\vartheta _{3}(2\pi
\theta ;q)\mathrm{d}\theta =q^{-n}(q,q)_{n}\delta _{nm}.  \label{a230}
\end{equation}
This relation is the well known orthogonality of the Rogers-Szeg\"{o}
polynomials on the unit circle. Clearly it is the same thing as the
orthogonality of the DG\ polynomials and one can be expressed in terms of
the other. This is also of interest because Macfarlane\cite{Macfarlane}
expressed orthogonality of his q-oscillator eigenfunctions in terms of
orthogonality of Rogers Szeg\"{o} polynomials on the unit circle.

\section{DG\ polynomials as a q-oscillator eigenfunctions}

\subsection{Algebraic derivation of DG\ polynomials}

Lets define the translation operator $T^{s}$ as $T^{s}=\mathrm{e}^{s\frac{%
\partial }{\partial x}}.$ It has the effect of shifting a function to the
left by $s$: $T^{s}f(x)=f(x+s).$ We define the creation and destruction
operators $\widehat{a}$ and $\widehat{a}^{\dagger }$ as
\begin{eqnarray}
\widehat{a} &=&\frac{1}{\sqrt{1-q}}T^{1/2}\left[ q^{x+1/4}-T^{1/2}\right] ,
\label{b10} \\
\widehat{a}^{\dagger } &=&\frac{1}{\sqrt{1-q}}\left[
q^{x+1/4}-T^{-1/2}\right] T^{-1/2}.  \label{b20}
\end{eqnarray}
Our inner product is the usual one
\begin{equation}
(f,g)=\int\limits_{-\infty }^{\infty }f^{*}(x)g(x)\mathrm{d}x,  \label{b30}
\end{equation}
and the conjugate operator is defined as $(f,\widehat{O}g)=(\widehat{O}%
^{\dagger }f,g).\,$According to this $(q^{x})^{\dagger }=q^{x}$ and $(\frac{%
\partial }{\partial x})^{\dagger }=-\frac{\partial }{\partial x}$ and $%
\widehat{a}^{\dagger }$ given above is the right one. The $\widehat{a}$ and $%
\widehat{a}^{\dagger }$ satisfy the commutation relation
\begin{equation}
\widehat{a}\widehat{a}^{\dagger }-q\widehat{a}^{\dagger }\widehat{a}=1.
\label{b40}
\end{equation}
This is the commutation relation satisfied by the Arik-Coon oscillator\cite
{Arik}.

We look for the eigenstates of the $\widehat{a}^{\dagger }\widehat{a}$
\begin{equation}
\widehat{a}^{\dagger }\widehat{a}A_{n}(x)=\lambda _{n}A_{n}(x).  \label{b50}
\end{equation}
We start from a 'ground state' $A_{0}(x)$ that satisfies $\widehat{a}%
A_{0}(x)=0$ which yields the functional equation
\begin{equation}
A_{0}(x+1/2)=q^{(x+1/4)}A_{0}(x).  \label{b70}
\end{equation}
If we put $A_{0}(x)=(cons.)w(x)q^{x^{2}}$ in this equation we get $%
w(x+1/2)=w(x).$ Therefore the normalized ground state is
\begin{equation}
A_{0}(x)=\alpha _{w}w(x)q^{x^{2}},  \label{b80}
\end{equation}
where $w(x)$ is any (in general complex) function satisfying $w(x+1/2)=w(x)$
periodicity condition and $\alpha _{w}$ is the normalization coefficient
\begin{equation}
\alpha _{w}=\left( \int\limits_{-\infty }^{\infty }\left| w(x)\right|
^{2}q^{2x^{2}}dx\right) ^{-1/2}.  \label{b90}
\end{equation}
For $w(x)=1$ we denote $\alpha _{w}$ as just $\alpha $ whose value is given
in eq. (\ref{a70}). We choose normalization of our eigenfunctions as $%
(A_{n},A_{n})=1$ and taking $w(x)=1$ with this normalization will lead us to
the normalized DG\ polynomials.

Next we build the states $(\widehat{a}^{\dagger })^{n}A_{0}(x)$. Using the
commutation relation one can easily show that if $(\widehat{a}^{\dagger
})^{n}A_{0}(x)$ is an eigenfunction of $\widehat{a}^{\dagger }\widehat{a}$
with the eigenvalue $\lambda _{n}$ then the $(\widehat{a}^{\dagger
})^{n+1}A_{0}(x)$ is an eigenfunction of the $\widehat{a}^{\dagger }\widehat{%
a}$ with eigenvalue $\lambda _{n+1}$ and one obtains a recursion relation
for the eigenvalues
\begin{equation}
\lambda _{n+1}=q\lambda _{n}+1.  \label{b60}
\end{equation}
Since $\widehat{a}A_{0}(x)=0$, then $A_{0}(x)$ is an eigenfunction of $%
\widehat{a}^{\dagger }\widehat{a}$ with the eigenvalue $\lambda _{0}=0$ and
by induction it follows that all $(a^{\dagger })^{n}A_{0}(x)$ are
eigenfunctions. Using the recursion relation in eq. (\ref{b60}) and $\lambda
_{0}=0$ we get the eigenvalues as
\begin{equation}
\lambda _{n}=\frac{1-q^{n}}{1-q}.  \label{b100}
\end{equation}

In exactly the same way that we do in solving the harmonic oscillator
algebraically, we can easily obtain the following relations
\begin{eqnarray}
\widehat{a}A_{n}(x) &=&\sqrt{\lambda _{n}}A_{n-1}(x),  \label{b110} \\
\widehat{a}^{\dagger }A_{n}(x) &=&\sqrt{\lambda _{n+1}}A_{n+1}(x).
\label{b120}
\end{eqnarray}
Then $A_{n}(x)$ can be written as
\begin{equation}
A_{n}(x)=\sqrt{\frac{(1-q)^{n}}{(q,q)_{n}}}(\widehat{a}^{\dagger
})^{n}A_{0}(x).  \label{b130}
\end{equation}
Instead of applying $\widehat{a}^{\dagger }$ n times, the following
recursive relation is easier. Define $A_{n}(x)$ as
\begin{equation}
A_{n}(x)=\frac{\alpha _{w}w(x)}{\sqrt{(q,q)_{n}}}%
\sum_{k=0}^{n}D_{k}^{n}(-1)^{k}q^{(n-k)/2}q^{(x-k)^{2}}.  \label{b140}
\end{equation}
Then if we apply $\widehat{a}^{\dagger }/\sqrt{\lambda _{n+1}}$ to obtain $%
A_{n+1}(x)$ and compare the coefficients we get the recursion relation for
the $D_{k}^{n}$ coefficients. In this process $w(x)$ completely commutes
with $\widehat{a}^{\dagger }$ because of periodicity: $%
T^{-1/2}w(x)=w(x-1/2)=w(x)$. The recursion relation we obtain is
\begin{equation}
D_{k}^{n+1}=q^{k}D_{k}^{n}+D_{k-1}^{n}.  \label{b150}
\end{equation}
Together with the conditions $D_{0}^{0}=1$, $D_{-1}^{0}=0$ this completely
determines the $D_{k}^{n}$. As can be shown easily, the q-binomial
coefficients $C_{k}^{n}$ satisfy this recursion relation and the boundary
conditions and therefore
\begin{equation}
D_{k}^{n}=C_{k}^{n}=\frac{(q,q)_{n}}{(q,q)_{k}(q,q)_{n-k}}.  \label{b160}
\end{equation}
This shows that
\begin{equation}
A_{n}(x)=\frac{\alpha _{w}w(x)}{\sqrt{(q,q)_{n}}}%
\sum_{k=0}^{n}C_{k}^{n}(-1)^{k}q^{(n-k)/2}q^{(x-k)^{2}},  \label{b170}
\end{equation}
are the normalized eigenfunctions and they reduce to the normalized DG\
polynomials $\phi _{n}(x)$ when $w(x)=1$.

Now let us see how the q-oscillator algebra reduces to harmonic oscillator
algebra in the limit $q\rightarrow 0$ $(c\rightarrow 0)$. The limit in
equation (\ref{a74}) shows us how to do it. First define the new variable $%
z=cx$. With this variable, the $\widehat{a}$ will look like
\begin{equation}
\widehat{a}=\frac{1}{\sqrt{1-\mathrm{e}^{-c^{2}}}}\mathrm{e}^{(c/2)\partial
/\partial z}\left[ \mathrm{e}^{-c^{2}/4}e^{-cz}-\mathrm{e}^{\frac{c}{2}%
(\partial /\partial z)}\right] ,  \label{b180}
\end{equation}
and in the limit $c\rightarrow 0$ this reduces to
\begin{equation}
\widehat{a}\rightarrow -\frac{1}{2}\frac{\partial }{\partial z}-z.
\label{b190}
\end{equation}
Similarly, $\widehat{a}^{\dagger }$ reduces to
\begin{equation}
\widehat{a}^{\dagger }\rightarrow \frac{1}{2}\frac{\partial }{\partial z}-z,
\label{b200}
\end{equation}
which, together with $\widehat{a}$, are the destruction and creation
operators for the harmonic oscillator problem
\begin{equation}
\left( -\frac{1}{4}\frac{\partial }{\partial z^{2}}+z^{2}\right) \psi
(z)=E\psi (z).  \label{b210}
\end{equation}

\subsection{Further discussion on $w(x)$ degeneracy}

Although the algebraic solution seems flawless, it is still very surprising
that the $A_{n}(x)$ are orthogonal for all the functions $w(x)$ satisfying $%
w(x+1/2)=w(x)$. Here we give a separate proof of it.

When we multiply two of our parent Gaussians $g_{n}(x)=q^{(x-n)^{2}}$ we get
daughter Gaussians $G_{n}(x)=q^{2(x-n/2)^{2}}$ as
\begin{equation}
g_{n}(x)g_{m}(x)=q^{(n-m)^{2}/2}G_{n+m}(x).  \label{c10}
\end{equation}
The parent Gaussians are centered at integers whereas the Daughter Gaussians
are centered at both integers and half integers. Therefore the $\phi
_{n}(x)\phi _{m}(x)$ product of the normalized DG\ polynomials
\begin{equation}
\phi _{n}(x)=\frac{\alpha }{\sqrt{(q,q)_{n}}}%
\sum_{k=0}^{n}C_{k}^{n}(-1)^{k}q^{(n-k)/2}q^{(x-k)^{2}},  \label{c20}
\end{equation}
can be written as a linear combination of Daughter Gaussians as
\begin{equation}
\phi _{n}^{*}(x)\phi _{m}(x)=\alpha
^{2}\sum_{k=0}^{n+m}d_{k}^{nm}q^{2(x-k/2)^{2}}.  \label{c30}
\end{equation}
If we integrate this we should get $\delta _{nm}$ due to the orthogonality
of normalized DG\ polynomials. The integrals $\int_{-\infty }^{\infty
}q^{2(x-k/2)^{2}}\mathrm{d}x$ are independent of $k$ (just shift the
integral by $k/2$) and have the value $1/\alpha ^{2}$. Therefore we get the
relation
\begin{equation}
\sum_{k=0}^{n+m}d_{k}^{nm}=\delta _{nm}.  \label{c40}
\end{equation}

Now, the $A_{n}^{*}(x)A_{m}(x)$ product of the $A_{n}(x)$ functions
\begin{equation}
A_{n}(x)=\frac{\alpha _{w}w(x)}{\sqrt{(q,q)_{n}}}%
\sum_{k=0}^{n}C_{k}^{n}(-1)^{k}q^{(n-k)/2}q^{(x-k)^{2}},  \label{c50}
\end{equation}
can be expressed as
\begin{equation}
A_{n}^{*}(x)A_{m}(x)=\left| \alpha _{w}\right|
^{2}\sum_{k=0}^{n+m}d_{k}^{nm}\left( \left| w(x)\right|
^{2}q^{2(x-k/2)^{2}}\right) .  \label{c60}
\end{equation}
If we integrate this we get
\begin{equation}
\int\limits_{-\infty }^{\infty }A_{n}^{*}(x)A_{m}(x)\mathrm{d}%
x=\sum_{k=0}^{n+m}d_{k}^{nm}\left( \left| \alpha _{w}\right|
^{2}\int\limits_{-\infty }^{\infty }\left| w(x)\right| ^{2}q^{2(x-k/2)^{2}}%
\mathrm{d}x\right) .  \label{c70}
\end{equation}
The integral in parenthesis can be shifted by $k/2$ as
\begin{equation}
\int\limits_{-\infty }^{\infty }\left| w(x)\right| ^{2}q^{2(x-k/2)^{2}}%
\mathrm{d}x=\int\limits_{-\infty }^{\infty }\left| w(x+k/2)\right|
^{2}q^{2x^{2}}\mathrm{d}x.  \label{c80}
\end{equation}
Because of the periodicity of $w(x)$, we have $w(x+k/2)=w(x)$ and all the
integrals are independent of $k$ and they have the value $1/\left| \alpha
_{w}\right| ^{2}$. This yields
\begin{equation}
\int\limits_{-\infty }^{\infty }A_{n}^{*}(x)A_{m}(x)\mathrm{d}%
x=\sum_{k=0}^{n+m}d_{k}^{nm}=\delta _{nm},  \label{c90}
\end{equation}
which follows from eq.(\ref{c40}). Therefore the orthogonality holds for any
$w(x)$ satisfying the periodicity requirement.

We actually found more than the DG polynomials from the algebraic treatment.
We found and infinite set of orthogonal functions.

\section{A different interpretation of q-oscillator example of Macfarlane}

In a seminal and widely cited paper Macfarlane constructed a different
coordinate representation of the q-oscillator. His definition of the
creation and destruction operators are
\begin{eqnarray}
\widehat{b} &=&\mathrm{e}^{2x}-\mathrm{e}^{x}\mathrm{e}^{s(\partial
/\partial x)},  \label{m10} \\
\widehat{b}^{\dagger } &=&\mathrm{e}^{-2x}-\mathrm{e}^{s(\partial /\partial
x)}\mathrm{e}^{-x}.  \label{m20}
\end{eqnarray}
Obviously, according to the usual definition of the inner product in eq.(\ref
{b30}) the $\widehat{b}^{\dagger }$ is not right. Macfarlane himself goes on
to construct eigenstates of $\widehat{b}^{\dagger }\widehat{b}$ without
discussing the inner product and orthogonality relation first. After finding
the eigenfunctions he states inner product and orthogonality relations in
terms of orthogonality of Rogers-Szeg\"{o} polynomials on the unit circle.
The functions we find do not agree with the functions he found and
apparently his formula for the eigenfunctions contains some error. Therefore
we redo the problem with a different inner product definition here.

We define the inner product as

\begin{eqnarray}
(f,g) &=&\int\limits_{-\infty }^{\infty }\left( \widehat{P}f^{*}(x)\right)
g(x)\mathrm{d}x.  \label{m30} \\
&=&\int\limits_{-\infty }^{\infty }f^{*}(-x)g(x)\mathrm{d}x  \label{m31}
\end{eqnarray}
where $\widehat{P}$ is the usual parity operator. For an operator $\widehat{O%
}$, its conjugate $\widehat{O}^{\dagger }$ is defined by the relation
\begin{equation}
(f,\widehat{O}g)=(\widehat{O}^{\dagger }f,g).  \label{m40}
\end{equation}
The $\widehat{b}^{\dagger }$ expression in eq.(\ref{m20}) is right if one
defines the inner product this way because $(\mathrm{e}^{x})^{\dagger }=%
\mathrm{e}^{-x}$ and $(\frac{d}{dx})^{\dagger }=\frac{d}{dx}$. Notice that
this definition of the inner product does not satisfy $(f,f)\geq 0$
condition. But we will not use this property of the inner product in our
development. We will use the conjugacy definition in eq.(\ref{m40}) in a few
places.

In order to obtain functions as a linear combination of the Gaussians
centered at nonnegative integers we change the variable $x=-c^{2}(y+1/4)$
and $s=-c^{2}/2$ and we take $q=\mathrm{e}^{-c^{2}}$ as before. We will also
divide $\widehat{b}$ and $\widehat{b}^{\dagger }$ by $\sqrt{q(1-q)}$ which
does not change the eigenfunctions (only eigenvalues), but in the limit of
the harmonic oscillator $(q\rightarrow 1)$ it helps to get the right
results. The new $\widehat{b}$ and $\widehat{b}^{\dagger }$ are
\begin{eqnarray}
\widehat{b} &=&\left( q^{2y+1/2}-q^{y+1/4}T^{1/2}\right) /\sqrt{q(1-q)},
\label{m50} \\
\widehat{b}^{\dagger } &=&\left( q^{-2y+1/2}-T^{1/2}q^{-y+1/4}\right) /\sqrt{%
q(1-q)},  \label{m60}
\end{eqnarray}
where $T^{1/2}=\mathrm{e}^{(\partial /\partial y)/2}$. They satisfy the
commutation relation
\begin{equation}
\widehat{b}^{\dagger }\widehat{b}-q\widehat{b}\widehat{b}^{\dagger }=1.
\label{m70}
\end{equation}
Note that this is somewhat different than the commutation relation in eq.(%
\ref{b40}).

Again we want to find the eigenfunctions of the operator $\ \widehat{b}%
^{\dagger }\widehat{b}$
\begin{equation}
\widehat{b}^{\dagger }\widehat{b}B_{n}(y)=\lambda _{n}B_{n}(y).  \label{m80}
\end{equation}
We again start from a ground state that satisfies $\widehat{b}B_{0}(y)=0$
(hence $\lambda _{0}=0$). $\widehat{b}B_{0}(y)=0$ yields the same functional
equation that $A_{0}(x)$ satisfy
\begin{equation}
B_{0}(y+1/2)=q^{y+1/4}B_{0}(y),  \label{m100}
\end{equation}
which we already know has the solution $B_{0}(y)=\alpha _{w}w(y)q^{y^{2}}$
where $w(y)$ satisfies the $w(y+1/2)=w(y)$ periodicity condition. The
infinite degeneracy of the states appear here too.

Next we build the unnormalized states $(\widehat{b}^{\dagger })^{n}B_{0}(y)$%
. Using the commutation relation we can easily show that if the $(\widehat{b}%
^{\dagger })^{n}B_{0}$ is an eigenfunction of $\widehat{b}^{\dagger }%
\widehat{b}$ with an eigenvalue $\lambda _{n}$ then $(\widehat{b}^{\dagger
})^{n+1}B_{0}$ is an eigenfunction with the eigenvalue $\lambda _{n+1}$
where $\lambda _{n+1}$ is related to the $\lambda _{n}$ as
\begin{equation}
q\lambda _{n+1}=\lambda _{n}-1.  \label{m90}
\end{equation}
Since $B_{0}(y)$ is an eigenfunction, by induction all $(\widehat{b}%
^{\dagger })^{n}B_{0}$ are eigenfunctions too. Starting from $\lambda _{0}=0$%
, this recursion relation yields
\begin{equation}
\lambda _{n}=-q^{-n}\left( \frac{1-q^{n}}{1-q}\right) .  \label{m110}
\end{equation}
Notice that the eigenvalues are negative.

To obtain relations similar to eqs.(\ref{b110},\ref{b120}) We form the inner
product
\begin{equation}
\lambda _{n}(B_{n},B_{n})=(B_{n},\widehat{b}^{\dagger }\widehat{b}B_{n})=(%
\widehat{b}B_{n},\widehat{b}B_{n}).  \label{m120}
\end{equation}
Taking $\widehat{b}B_{n}=\mu _{n}B_{n-1}$ we get
\begin{equation}
\lambda _{n}=\left| \mu _{n}\right| ^{2}\frac{(B_{n-1},B_{n-1})}{%
(B_{n},B_{n})}.  \label{m130}
\end{equation}
Since $\left| \mu _{n}\right| ^{2}$ is positive and $\lambda _{n}$ is
negative, the $(B_{n-1},B_{n-1})$ and $(B_{n},B_{n})\,$must have opposite
signs. This looks surprising but we should remember that in our definition
of the inner product the $(f,f)$ can be negative. Therefore we will take our
normalization as
\begin{equation}
(B_{n},B_{n})=(-1)^{n}.  \label{m140}
\end{equation}
With this normalization we obtain $\mu _{n}=\sqrt{-\lambda _{n}}$:
\begin{equation}
\widehat{b}B_{n}=\sqrt{-\lambda _{n}}B_{n-1}.  \label{m150}
\end{equation}
Starting from
\begin{equation}
\lambda _{n+1}B_{n+1}=\widehat{b}^{\dagger }\widehat{b}B_{n+1}=\sqrt{%
-\lambda _{n+1}}\widehat{b}^{\dagger }B_{n},  \label{m160}
\end{equation}
we also obtain
\begin{equation}
\widehat{b}^{\dagger }B_{n}=-\sqrt{-\lambda _{n+1}}B_{n+1}.  \label{m170}
\end{equation}
This result is also a little unusual because of the sign in front. It is a
consequence of negative $\lambda _{n}$ eigenvalues.

Since we started from an unusual inner product definition there might be
doubts on orthogonality of the $B_{n}$. Consider the inner product
\begin{equation}
(B_{m},\widehat{b}^{\dagger }\widehat{b}B_{n})=(\widehat{b}^{\dagger }%
\widehat{b}B_{m},B_{n}).  \label{m180}
\end{equation}
Using $\widehat{b}^{\dagger }\widehat{b}B_{k}=\lambda _{k}B_{k}$ we get
\begin{equation}
(\lambda _{n}-\lambda _{m})(B_{m},B_{n})=0,  \label{m190}
\end{equation}
which yields $(B_{m},B_{n})=0$ when $m\neq n$. We added this common proof to
emphasize that the orthogonality does not depend on the $(f,f)\geq 0$
property of the usual inner products. We just used the conjugacy relation
eq.(\ref{m40}) in this proof of orthogonality. But unlike the case of usual
definition inner product given in eq.(\ref{b30}), $(f,f)=0$ does not imply $%
f=0$ when we don't have $(f,f)\geq 0$ property.

To complete the discussion we obtain the eigenfunctions $B_{n}(y)$. It makes
things easier to figure out the coefficient of Gaussian centered at zero($%
q^{y^{2}}$) first. The $\widehat{b}^{\dagger }$
\begin{equation}
\widehat{b}^{\dagger }=\left( q^{-2y+1/2}-T^{1/2}q^{-y+1/4}\right) /\sqrt{%
q(1-q)},  \label{m192}
\end{equation}
has two parts that do different things. When $q^{-2y+1/2}$ acts on $%
q^{(y-k)^{2}}$ it produces the next Gaussian $q^{(y-k-1)^{2}}$ and
multiplies it with a constant. When $T^{1/2}q^{-y+1/4}$ act on $%
q^{(y-k)^{2}} $ it gives $q^{(y-k)^{2}}$ back and multiplies with a
constant. Here are the precise relations
\begin{eqnarray}
q^{-2y+1/2}q^{(y-k)^{2}} &=&q^{-2k-1/2}q^{(y-k-1)^{2}},  \label{m194} \\
T^{1/2}q^{-y+1/4}q^{(y-k)^{2}} &=&q^{-k}q^{(y-k)^{2}}.  \label{m196}
\end{eqnarray}
From the second relation we have $T^{1/2}q^{-y+1/4}q^{y^{2}}=q^{y^{2}}$
which means this operator leaves $q^{y^{2}}$ as it is. The other operator $%
q^{-2y+1/2}\,$creates $k=1$ Gaussian ($q^{(y-1)^{2}}$) from it. Since we
produce higher eigenfunctions by applying $-\widehat{b}^{\dagger }/\sqrt{%
-\lambda _{n}}$ successively, after each application the coefficient of $k=0$
Gaussian changes by a factor $1/\sqrt{q(1-q)(-\lambda _{n})}$. We denote the
coefficient of $k=0$ Gaussian in $B_{n}(y)$ by $\zeta _{n}$ and it should be
\begin{equation}
\zeta _{n}=\alpha _{w}\frac{1}{\sqrt{q^{n}(1-q)^{n}}}\frac{1}{\sqrt{%
(-\lambda _{1})(-\lambda _{2})...(-\lambda _{n})}}=\alpha _{w}\frac{%
q^{n(n-1)/4}}{\sqrt{(q,q)_{n}}}.  \label{m198}
\end{equation}

Now let us  take the $B_{n}(y)$ of the form
\begin{equation}
B_{n}(y)=w(x)\zeta _{n}\sum\limits_{k=0}^{n}E_{k}^{n}q^{(y-k)^{2}},
\label{m201}
\end{equation}
where $E_{0}^{n}=1$ by construction. We can use Eq. (\ref{m170}) to generate
a recursion relation for $E_{k}^{n}$ as before. To show a different and
easier way of doing things we will use eq.(\ref{m150}) relation this time.
The $\widehat{b}$ has the effect of shifting each Gaussian to the left by
one unit
\begin{equation}
\widehat{b}q^{(y-k)^{2}}=-\frac{q^{k-1/2}(1-q^{k})}{\sqrt{q(1-q)}}%
q^{(y-k+1)^{2}},  \label{m209}
\end{equation}
and it also destroys the leftmost Gaussian (the $k=0$ Gaussian centered at
zero). Therefore the equality
\begin{equation}
\widehat{b}\left[ \left( \zeta _{n}E_{k}^{n}\right) q^{(y-k)^{2}}\right] =%
\sqrt{-\lambda _{n}}\left[ \left( \zeta _{n-1}E_{k-1}^{n-1}\right)
q^{(y-k+1)^{2}}\right]   \label{m212}
\end{equation}
should hold for each Gaussian. This yields the recursion relation for the $%
E_{k}^{n}$ as
\begin{equation}
E_{k}^{n}=-E_{k-1}^{n-1}\left( \frac{1-q^{n}}{1-q^{k}}\right) q^{-n-k+3/2}.
\label{m220}
\end{equation}
Together with the condition $E_{0}^{n}=1$ this is enough information to
solve the $E_{k}^{n}$. The quotient in parenthesis tells us that the q
binomial coefficients $C_{k}^{n}$ are involved. If we set $%
E_{k}^{n}=(-1)^{k}C_{k}^{n}q^{u(n,k)}$ we get the recursion relation for $%
u(n,k)$ as
\begin{equation}
u(n,k)-u(n-1,k-1)=-n-k+3/2.  \label{m230}
\end{equation}
Together with the condition $u(n,0)=0$ (which follows from $E_{0}^{n}=1$)$\,$%
this is uniquely solved as $u(n,k)=-nk+k/2$. Therefore the $B_{n}(y)$ should
be
\begin{equation}
B_{n}(y)=\alpha _{w}w(x)\frac{q^{n(n-1)/4}}{\sqrt{(q,q)_{n}}}%
\sum\limits_{k=0}^{n}C_{k}^{n}(-1)^{k}q^{-(n-1/2)k}q^{(y-k)^{2}}.
\label{m240}
\end{equation}

This formula does not agree with the result of Macfarlane even after setting
$c^{2}=-2s$ and $y=-1/4+x/2s$ back in $B_{n}(y)$ above. To be sure that we
have the right formula we have checked numerically if the $B_{n}(y)$ satisfy
the orthogonality relation

\begin{equation}
\int\limits_{-\infty }^{\infty }B_{n}(-y)B_{m}(y)\mathrm{d}y=(-1)^{n}\delta
_{nm}.  \label{m250}
\end{equation}
They satisfy it perfectly and we are sure that we have the right formula.
Apparently Macfarlane's paper contains an error.

The harmonic oscillator limit is as straightforward as it is in DG\
polynomials case. We change variable $y=z/c$ and take the limit $%
c\rightarrow 0$ which yields
\begin{eqnarray}
\widehat{b} &\rightarrow &z-\frac{1}{2}\frac{\partial }{\partial z},
\label{m253} \\
\widehat{b}^{\dagger } &\rightarrow &-z-\frac{1}{2}\frac{\partial }{\partial
z}.  \label{m354}
\end{eqnarray}
Then $\widehat{b}^{\dagger }\widehat{b}B_{n}=\lambda _{n}B_{n}$ reduces to ($%
\lambda _{n}\rightarrow -n$ in this limit)
\begin{equation}
(-\frac{1}{4}\frac{\partial ^{2}}{\partial z^{2}}+z^{2})\Psi
_{n}=(n+1/2)\Psi _{n}.  \label{m256}
\end{equation}
The fact that the inner product is defined differently makes no difference
in this limit because the harmonic oscillator eigenfunctions are either even
or odd. The harmonic oscillator eigenfunctions satisfy the normalization
condition
\begin{equation}
\int \Psi _{n}(-y)\Psi _{m}(y)\mathrm{d}y=(-1)^{n}\delta _{nm}  \label{m258}
\end{equation}
as can be verified easily using parity of the wave functions.

Now, just as the DG\ polynomials, the orthogonality of the $B_{n}(y)$ can be
expressed as an orthogonality relation of the Rogers-Szeg\"{o} polynomials
on the unit circle. We will take $w(x)=1$ and $\alpha _{w}=\alpha $ for
this. By Fourier transforming the orthogonality relation eq.(\ref{m250}) we
get the relation
\begin{equation}
\int\limits_{-\infty }^{\infty }B_{n}^{*}(-\theta )B_{m}(\theta )\mathrm{d}%
\theta =\ (-1)^{n}\delta _{nm},  \label{m260}
\end{equation}
where $B_{n}(\theta )$ is the Fourier transform of the $B_{n}(y)$:
\begin{equation}
B_{n}(\theta )=\left( \frac{\pi }{c^{2}}\right) ^{1/2}\zeta _{n}\mathrm{e}%
^{-(\pi /c)^{2}\theta ^{2}}\sum_{k=0}^{n}C_{k}^{n}(-q^{-(n-1/2)}\mathrm{e}^{%
\mathrm{i}2\pi \theta })^{k}.  \label{m280}
\end{equation}
Here $\theta $ is the Fourier transform variable just as before. Using the
Poisson summation formula in eq. (\ref{a190}) the orthogonality relation can
be expressed as
\begin{equation}
\int\limits_{0}^{1}H_{n}(-q^{-(n-1/2)}\mathrm{e}^{\mathrm{i}2\pi \theta
})H_{m}(-q^{-(m-1/2)}\mathrm{e}^{\mathrm{i}2\pi \theta })\left(
\sum\limits_{k=-\infty }^{\infty }\mathrm{e}^{-2(\pi /c)^{2}(\theta
+k)^{2}}\right) \mathrm{d}\theta =\ \frac{c^{2}}{\pi }\frac{(-1)^{n}}{\zeta
_{n}^{2}}\delta _{nm}.  \label{m290}
\end{equation}
Setting the sum in parenthesis from the eq. (\ref{a210}) and (\ref{a220}) we
obtain
\begin{equation}
\int\limits_{0}^{1}H_{n}(-q^{-(n-1/2)}\mathrm{e}^{\mathrm{i}2\pi \theta
})H_{m}(-q^{-(m-1/2)}\mathrm{e}^{\mathrm{i}2\pi \theta })\vartheta _{3}(2\pi
\theta ;q)\mathrm{d}\theta =\ q^{-n(n-1)/2}(q,q)_{n}(-1)^{n}\delta _{nm}.
\label{m300}
\end{equation}
This is again some form of the orthogonality relation of the Rogers-Szeg\={o}
polynomials on the circle. This is a new set of orthogonality relations and
we are not aware of its existence in mathematical literature.

\section{Summary}

In this study we found that the DG\ polynomials that Karabulut and Sibert
discovered before are actually eigenfunctions of coordinate representation
of the Arik-Coon q-oscillator. We derived the DG\ polynomials from
q-oscillator algebra. We also indicated that orthogonality of the DG\
polynomials can be cast into orthogonality of the Rogers-Szeg\"{o}
polynomials on the unit circle and the two are equivalent.

We showed that the example given by Macfarlane can be interpreted with an
unusual inner product definition and we constructed the corresponding
orthogonal functions. Their orthogonality can be recast into a form of
orthogonality relation for the Rogers Szeg\"{o} polynomials on the unit
circle. We were not able to find this result in mathematical literature and
it is probably a new result.

A very interesting result of this work is that the eigenstates of the
q-oscillators we solved turned out to be infinitely degenerate. We know that
algebraic solution of the one dimensional harmonic oscillator is
nondegenerate. In the usual algebraic solution of the harmonic oscillator we
base our arguments on the commutation relations and the algebra has nothing
in it that implies nondegenerate states. But the ground state $\widehat{a}%
\phi _{0}=0$ yields a unique solution for the harmonic oscillator because it
is a differential equation and nondegeneracy of the excited states follows
from this. For our q-oscillator examples we have a first order difference
equation for the $\widehat{a}\phi _{0}=0$ and such equations together with
the boundary conditions do not uniquely define a function. It defines a
function on all real axis if  its values in a 1/2 wide interval are known.

Finally, the freedom to choose $w(x)$ arbitrarily gives us possibility of
constructing orthogonal function sets more general than the DG polynomials.
Consider a set of functions $w_{0}(x),w_{1}(x),w_{2}(x)....$ satisfying the
periodicity condition $w_{n}(x+1/2)=w_{n}(x)$ $(n=0,1,2,...)$ and the
orthogonality relation
\begin{equation}
\int\limits_{-\infty }^{\infty }w_{n}^{*}(x)w_{m}(x)q^{2x^{2}}\mathrm{d}%
x=\delta _{nm}.  \label{s10}
\end{equation}
Then the set of functions
\begin{equation}
\Gamma _{nm}(x)=\frac{\alpha (w_{n})}{\alpha }w_{n}(x)\phi _{m}(x)
\label{s15}
\end{equation}
will satisfy an orthogonality relation of the form
\begin{equation}
\int\limits_{-\infty }^{\infty }\Gamma _{nm}^{*}(x)\Gamma _{ij}(x)\mathrm{d}%
x=\delta _{ni}\delta _{mj}.  \label{s20}
\end{equation}
Here $\alpha (w_{n})$ is the $\alpha _{w}$ for $w=w_{n}(x)$ given in eq.(\ref
{b90}) and $\alpha $ is given in eq.(\ref{a70}). This gives us much freedom
to construct orthogonal function sets useful as basis sets in variational
calculations. This possibility should be investigated in future research.

\noindent\textbf{Acknowledgments}

In a private communication Prof. R. G. Littlejohn suggested to the author
that there must be an operator that admits the DG\ polynomials as
eigenfunctions, or equivalently, there must be an algebraic structure behind
them. Although his insightful comments and suggestions are entirely
responsible for the creation of the work presented here, he kindly declined
to coauthor this paper on the grounds that his contribution was minimal. {%
The author is indebted to Professor Robert G. Littlejohn for suggesting the
problem addressed in this paper.}

\end{document}